\documentclass[referee]{aa} 

\usepackage{graphicx,epstopdf}
\usepackage{txfonts}

\begin{document}

   \title{Toward a fast and consistent approach to model solar magnetic fields in multiple layers}

   \author{X. Zhu\inst{1} and T. Wiegelmann\inst{2}}

  \institute{Key Laboratory of Solar Activity, National Astronomical Observatories, Chinese Academy of Sciences, Beijing, China\\
            \email{xszhu@bao.ac.cn}
            \and
            Max-Planck-Institut f\"{u}r Sonnensystemforschung, Justus-von-Liebig-Weg 3, 37077 G\"{o}ttingen, Germany}

   \date{Received ; accepted }

  \abstract
   {}
   {We aim to develop a fast and consistent extrapolation method to model multiple layers of the solar atmosphere.}
   {The new approach combines the magnetohydrostatic (MHS) extrapolation which models the solar low atmosphere in a flat box and the nonlinear force-free field (NLFFF) extrapolation which models the solar corona with a chromospheric vector magnetogram deduced from the MHS extrapolation. We test our code with a snapshot of a radiative magnetohydrodynamic simulation of a solar flare. Comparisons are conducted by several metrics quantitatively.}
   {Based on a number of test runs, we find out the optimized configuration for the combination of two extrapolations with a 5.8-Mm-high box for the MHS extrapolation and a magnetogram at a height of 1 Mm for the NLFFF extrapolation. The new approach under this configuration is able to reconstruct the magnetic fields in multi-layers accurately and efficiently. Based on figures of merit that are used to assess the performance of different extrapolations (NLFFF extrapolation, MHS extrapolation, and the combined one), we find the combined extrapolation reaches accuracy of the MHS extrapolation which is better than the NLFFF extrapolation. The combined extrapolation is moderately efficient for application to magnetograms with high resolution.}
   {}


\titlerunning{A consistent model for the solar magnetic field}
\maketitle

\section{Introduction}


Over half a century, numerous extrapolations have been developed to compute magnetic fields from the photosphere to the corona \citep[see reviews by][]{r13,ws21}. As one of the most used extrapolation methods in modeling coronal magnetic fields of an active region, a nonlinear force-free field (NLFFF) extrapolation assumes a vanishing Lorentz force in the solar corona in which the plasma $\beta$ is very low. Then the coronal magnetic fields can be described by
\begin{eqnarray}
\nabla\times {\bf B}&=&\alpha {\bf B}, \label{eq:ff}\\
\nabla\cdot {\bf B}&=&0.\label{eq:divb1}
\end{eqnarray}
Take the divergence of Eq.~(\ref{eq:ff}) and use Eq.~(\ref{eq:divb1}), we get
\begin{equation}
{\bf B}\cdot\nabla\alpha=0 \label{eq:alpha}
\end{equation}
showing $\alpha$ is constant along each magnetic field line but can have different values on different field lines. In lower atmosphere, however, magnetic field is not necessarily force-free due to non-negligible gas pressure gradient and gravity. The non-force-freeness of the magnetic field in the lower atmosphere were confirmed by a couple of studies \citep{mjm95,mcy02,t12,lsz13,lh15} which showed that most photospheric magnetograms of active regions fail to fulfill the Aly's criteria \citep{m69,m74,a84,a89} for a force-free field. The inconsistency between the force-free assumption and the non-force-free magnetogram has prevented NLFFF extrapolations from acquiring good results \citep{mds08,dsb09}.

To overcome the inconsistency problem one may perform the NLFFF extrapolation by using a chromospheric magnetogram. It was found that the magnetic field becomes force-free roughly 400 km above the photosphere \citep{mjm95}. However, the chromosphere is thicker than the photosphere and chromospheric spectrum lines are formed in a wide range of heights \citep{blr19} which make it difficult to tell at which height a magnetogram is located. Moreover, interpreting spectropolarimetric measurements made using chromospheric spectrum lines is still a challenge \citep{h12}. Another way to deal with the inconsistency is to preprocess the photospheric magnetogram before it is applied. \cite{wis06} first proposed an algorithm to modify magnetogram within the error margins of measurement to minimize the net magnetic force and torque in the data. After that, several other preprocessing methods have been developed \citep{fsv07,twi09,yk12,jf14}. The preprocessed magnetogram is much more suitable for a NLFFF extrapolation than the unpreprocessed one. The third way which seems more natural is to design the magnetohydrostatic (MHS) extrapolation to solve the following MHS equations
\begin{eqnarray}
\frac{1}{4\pi}(\nabla \times \mathbf{B})\times \mathbf{B}-\nabla p - \rho {\bf g}\mathbf{\hat{z}} & = & 0, \label{eq:force_balance}\\
\nabla \cdot \mathbf{B} & = & 0, \label{eq:divB}
\end{eqnarray}
where $\mathbf{B}$, $p$, $\rho$, and $\bf g$ are the magnetic field, plasma pressure, plasma density, and gravitational acceleration, respectively. The magnetic fields and plasma interact consistently in the MHS extrapolation, which allows the photospheric magnetogram to not fulfill the Aly's criteria. The MHS extrapolations have been developed based on magnetohydrodynamic relaxation method \citep{zwd13,zwd16,mki20}, \cite{gr58} method \citep{gw13,gbb16}, and optimization method \citep{wn06,wnr07,zw18,zw19}. Applications of the MHS extrapolation mainly focus on activities in the low atmosphere \citep{zsl17,tzp18,ctz19,stz20,zwd16,zwc17,zws20,jwf21}.

It is worth noting that a new class of the NLFFF extrapolations by using multiple observations has been introduced recently \citep{msd12,a13,ciw15,dsg19}. Besides the photospheric magnetogram, these extrapolations take advantage of either EUV images \citep[AIA,][]{lta12} or coronal polarimetric observations \citep[CoMP,][]{tcd08}. As we getting more and more data such as polarimetric observation with various wavelengths by DKIST \citep{trb16} and radio observation with various frequencies by MUSER \citep{yzw09,ycw21}, etc, which are sensitive to the magnetic fields, it is a promising way to optimize extrapolation with the multiple observations.

In this work we try to address two issues with the first one to validate the effectiveness of the MHS extrapolation applied to the solar corona and the second one to improve the efficiency of the MHS extrapolation which has been found rather time-consuming. The reference solution is a snapshot of a radiative MHD (RMHD) simulation of a solar flare, which has been used to test the MHS extrapolation in the lower solar atmosphere \citep{zw19}. The remainder of the paper is organized as follows: The reference solution is described in Sect.~\ref{sec:simulation}. The application of the MHS extrapolation to the corona is tested in Sect.~\ref{sec:mhs}. Incorporation of the two extrapolations are tested in Sect.~\ref{sec:mhsnlff}. Conclusions are presented in Sect.~\ref{sec:conclusion}.

\section{Active region in an RMHD simulation}\label{sec:simulation}

\cite{crc19} carried out an RMHD simulation of a solar flare along with a coronal mass ejection. The RMHD code includes 3D radiative transfer in the convection zone and photosphere, optically thin radiation and field-aligned heat conduction in the corona, which mimics a realistic environment on the Sun. The eruption of the simulation was triggered by imposing an emergence of a twisted flux rope into a pair of initially potential sunspots.

\begin{figure*}
  \centering
  \includegraphics[width=17cm]{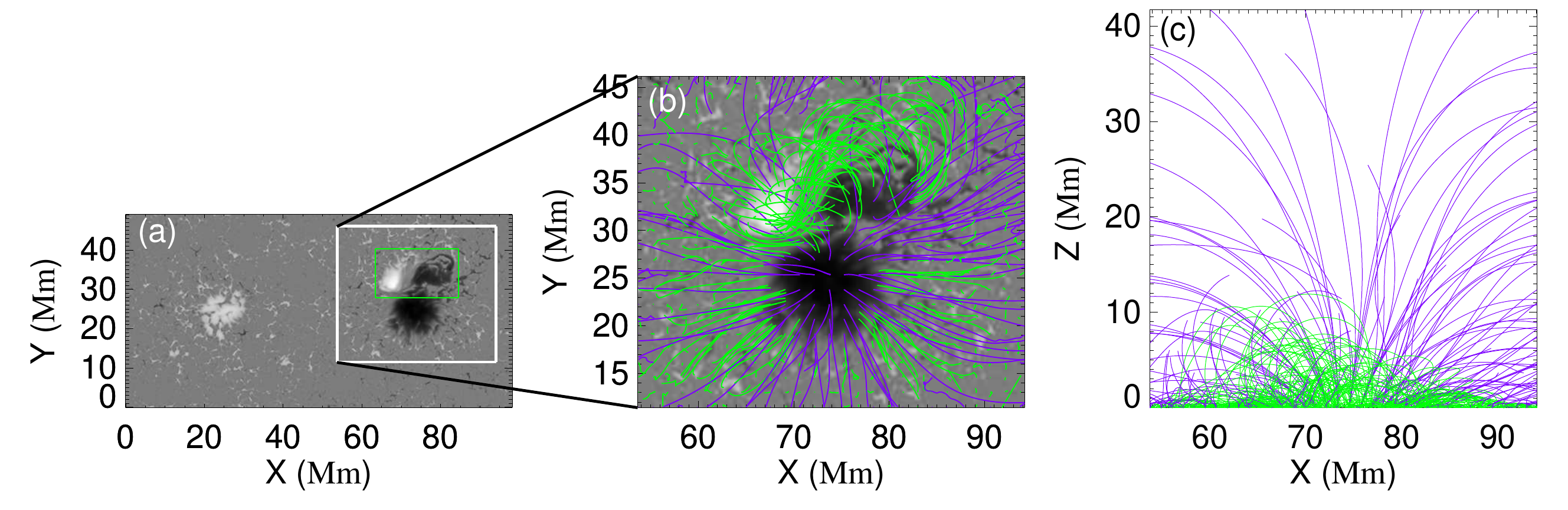}
  \caption{(a): Photospheric magnetogram of the reference model. (b): Sample magnetic field lines in the white box outlined in panel (a). Purple/green lines are open/closed magnetic field lines. (c) The same field lines seen from south.}
  \label{fig:ref_model}
\end{figure*}

We chose a snapshot that was 8 minutes after flare peak as the reference model. The reference model was previously used to test the MHS extrapolation in the lower atmosphere \citep{zw19}. The original data spans $\pm49.2 (\pm24.6 Mm)$ in x (y) axis and spans from 7.5 Mm beneath the photosphere to 41.6 Mm above in z axis. The grid resolution is 192 km in the horizontal directions and 64 km in the vertical. The photospheric magnetogram (Fig.~\ref{fig:ref_model} (a)) is located at the average geometrical height corresponding to optical depth unity. Fig.~\ref{fig:ref_model} (b) and (c) show magnetic field lines from different perspectives. We see that the top of the flux rope is roughly 10 Mm in the low solar corona, which is much higher than chromospheric structures that typically have a height of 2 Mm.

\begin{figure}
  \resizebox{\hsize}{!}{\includegraphics{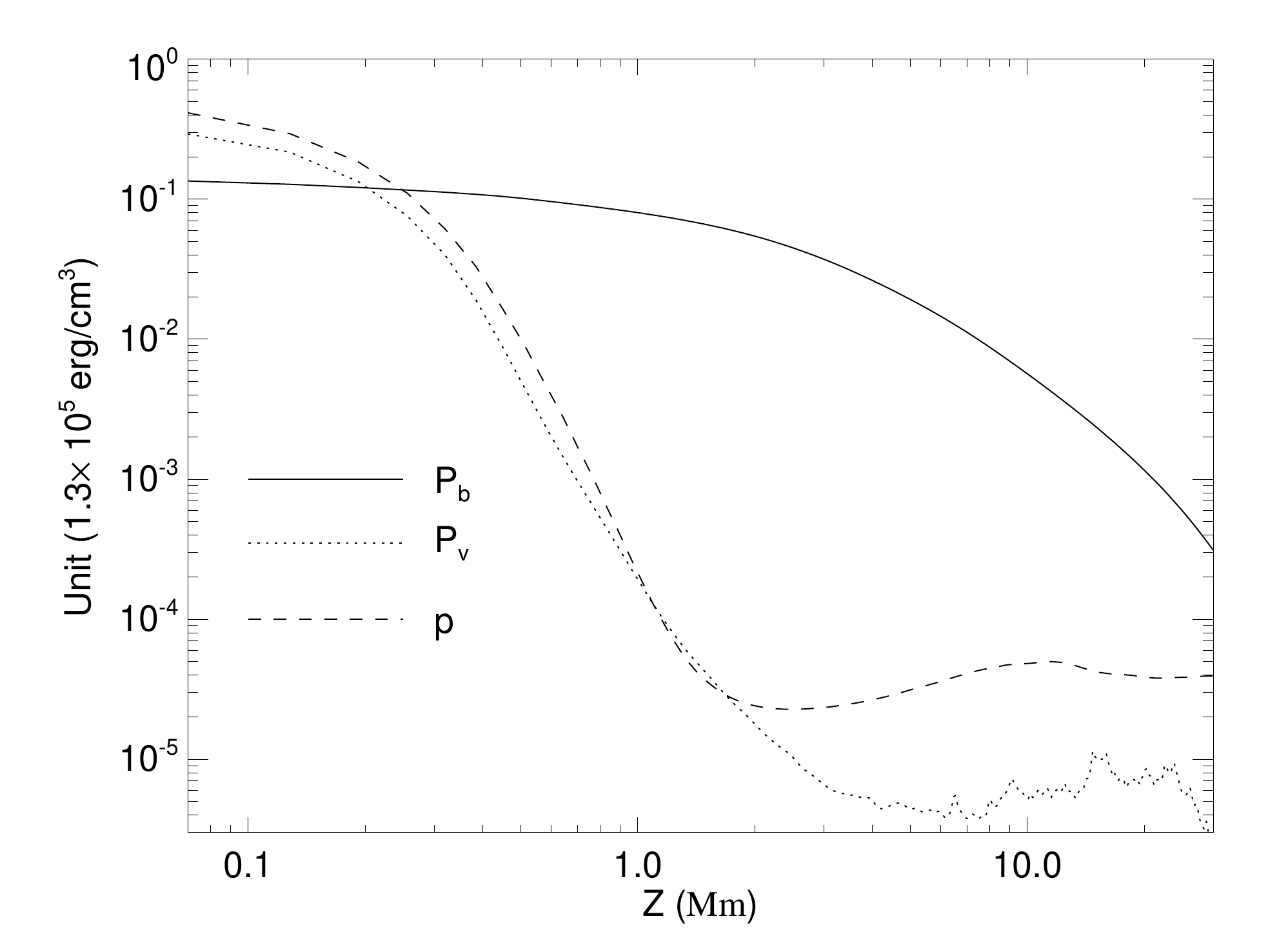}}
  \caption{Horizontal averages of the reference model. $P_{b}$ (solid line), $P_{v}$ (dotted line), and p (dashed line) are horizontally averaged magnetic pressure, inertial term, and plasma pressure, respectively.}
  \label{fig:pbpgpv}
\end{figure}

Although the system was still dynamic at the moment of the reference snapshot, we found out that the time-varying term is at least 1 order smaller than other terms in the moment equation (see Fig.1 (a) of \cite{zw19}). Fig.~\ref{fig:pbpgpv} shows influence of magnetic pressure ($P_b=B^2/2$), plasma pressure ($p$), and inertial term ($P_v=\rho v^2$) at different layers. We see the plasma pressure is the most important term in the very low atmosphere (below $250$ km), while the magnetic term dominates the higher layers. Recall that the MHS model does not consider the inertial force. Hence the snapshot of the RMHD simulation serves an adequate reference model to test our code.

\begin{figure}
  \resizebox{\hsize}{!}{\includegraphics{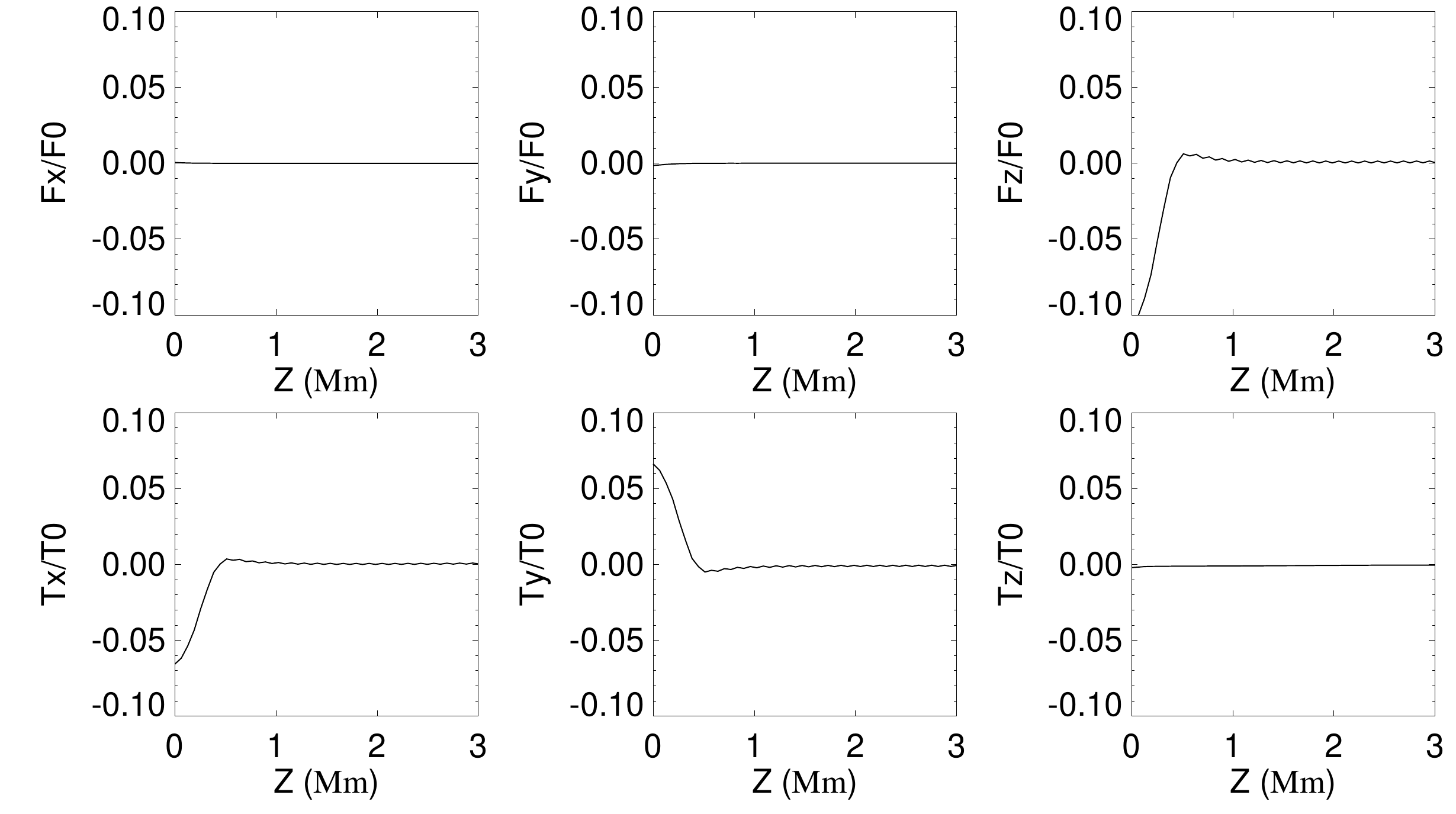}}
  \caption{Scaled net Lorentz force and torque with $F_{0}=\int_{S_1} ({B_x}^2+{B_y}^2+{B_z}^2)dxdy$ and $T_{0}=\int_{S_1}\sqrt{x^2+y^2}({B_x}^2+{B_y}^2+{B_z}^2)dxdy$ as functions of height.}
  \label{fig:net_force_torque}
\end{figure}

We also checked the magnetogram to see to what degree it satisfies with criteria for either a force-free or a MHS state. The Aly critera \citep{m69,m74,a84,a89} which consist of 6 integrals on the magnetogram are necessary conditions for the magnetic field to be force-free. They are given by
\begin{eqnarray}
F_{x}&=-&\int_{S_1} B_xB_zdxdy = 0\label{eq:fx}, \\
F_{y}&=-&\int_{S_1} B_yB_zdxdy = 0\label{eq:fy}, \\
F_{z}&=&\int_{S_1} \left(\frac{{B_x}^2+{B_y}^2-{B_z}^2}{2}\right)dxdy = 0\label{eq:fz}, \\
T_{x}&=&\int_{S_1} y\left(\frac{{B_x}^2+{B_y}^2-{B_z}^2}{2}\right)dxdy = 0\label{eq:tx}, \\
T_{y}&=-&\int_{S_1} x\left(\frac{{B_x}^2+{B_y}^2-{B_z}^2}{2}\right)dxdy = 0\label{eq:ty}, \\
T_{z}&=&\int_{S_1} (yB_zB_x-xB_zB_y)dxdy = 0\label{eq:tz},
\end{eqnarray}
where $S_1$ is the bottom boundary. Equations (\ref{eq:fx}-\ref{eq:fz}) represent the null net Lorentz force at each of three directions, while equations (\ref{eq:tx}-\ref{eq:tz}) represent the null net Lorentz force torque at each of three directions. The necessary conditions for the magnetic field to be in a MHS state are weaker because of the existence of plasma \citep{zwi20}, which only consist of the equations (\ref{eq:fx})(\ref{eq:fy})(\ref{eq:tz}). Moreover, based on the equations (\ref{eq:fx})(\ref{eq:fy})(\ref{eq:tz}), \cite{zwi20} developed a preprocessing procedure to remove the net transverse forces and vertical torque on magnetogram while the net vertical force and transverse torques remain. The preprocessed magnetogram is consistent with the MHS assumption, and hence is supposed to be more appropriate for a MHS extrapolation. Fig.~\ref{fig:net_force_torque} shows changes of scaled net Lorentz force and torque along height. We can see $F_{x}/F_{0}$, $F_{y}/F_{0}$, $T_{z}/T_{0}$ are very close to zero in all layers while $F_{z}/F_{0}$, $T_{x}/T_{0}$, $T_{y}/T_{0}$ are around 0.1 in the photosphere, which means the equations (\ref{eq:fx})(\ref{eq:fy})(\ref{eq:tz}) are satisfied well while the vertical force and transverse torques are somewhat large. Therefore the magnetogram may be used in the MHS extrapolation directly but should be preprocessed before being used in the NLFFF extrapolation. It is also worth to note that $F_{z}/F_{0}$, $T_{x}/T_{0}$, $T_{y}/T_{0}$ decrease quickly and almost vanish above 400 km.


\section{MHS extrapolation testing in the solar corona}\label{sec:mhs}
\subsection{Methodology and numerical setting}
To solve the MHS equations (\ref{eq:force_balance}-\ref{eq:divB}), we construct the functional
\begin{equation}
\begin{aligned}
L(\mathbf{B},p,\rho)=\int_{V}\omega_{a}B^{2}\Omega_{a}^{2}+\omega_{b}B^{2}\Omega_{b}^{2}dV +\nu \int_{S}(\mathbf{B}-\mathbf{B}_{\!o\;\!\!b\;\!\!s})\cdot\mathbf{W}\cdot(\mathbf{B}-\mathbf{B}_{\!o\;\!\!b\;\!\!s})dS,
\end{aligned}
\label{eq:L}
\end{equation}
with
\begin{eqnarray}
\mathbf{\Omega_{a}} &=& \left[(\nabla \times \mathbf{B})\times \mathbf{B}-\nabla p - \rho \mathbf{\hat{z}}\right]/(B^2+p), \label{eq:Omga}\\
\mathbf{\Omega_{b}} &=& [(\nabla \cdot \mathbf{B})\mathbf{B}]/(B^2+p),\label{eq:Omgb}
\end{eqnarray}
where $\omega_{a}$ and $\omega_{b}$ are the weighting functions which are set to unity over the whole region in this study, $\nu$ is the Lagrangian multiplier. $\mathbf{B}_{obs}$ are the observed magnetic fields in the photosphere. W is a diagonal matrix to incorporate measurement error of the magnetic field. We set the vertical component $w_{los}$ to unity, while the transverse component $w_{tran}=\frac{B_{T}}{max(B_{T})}$. It is obvious that equations (\ref{eq:force_balance}-\ref{eq:divB}) are fulfilled when $L=0$. By using the gradient descent method the functional L is minimized. Note that, to make the formulae (\ref{eq:Omga}) simplified, coefficients $\frac{1}{4\pi}$ and $\bf g$ disappear due to appropriate normalization of physical quantities. For example, for a study that includes photosphere, the following normalization constants are convenient: $\rho_0=2.7\times10^{-1}g/cm^3$ (density), $T_0=6\times 10^3K$ (temperature), $g=2.7\times10^4 cm/s^2$ (gravitational acceleration), $L_0=\frac{\mathcal{R}T_0}{\mu g}=1.8\times10^7 cm$ (length), $p_0=\sqrt{\frac{\rho_0\mathcal{R}T_0}{\mu}}=1.3\times10^5 dyn/cm^2$ (plasma pressure), and $B_0=\sqrt{4\pi p_0}=1.3\times10^3 G$ (magnetic field), where $\mathcal{R}$ is the ideal gas constant.

The extrapolation follows the same dimensions of the reference model with $512\times256\times652$ grid points in the (x, y, z) directions and with a grid spacing of 192 km horizontally and 64 km vertically. In a previous study \citep{zw19} we used a flat box with $98\times49\times8$ $Mm^{3}$ to test the code in the chromosphere. The computational box is extended to the corona in this study.

As the initial condition we compute a NLFFF by an optimization method \citep{w04} with the preprocessed magnetogram \citep{wis06}. During the NLFFF step the preprocessed magnetogram is injected slowly \citep{wi10}. By slow injection the potential field at the boundary adjusted continuously according to the gradient descent method over the total iterations of the calculation (usually a few thousand steps). Then we move to the next step of taking into account the effects of plasma forces. In this step, however, the original magnetogram is injected slowly. The side and upper boundaries of the magnetic field and the plasma are fixed to their initial values. The iteration stops when $\frac{dL}{dt}/L < 10^{-3}$ for 500 consecutive steps. For more details of the implementation of the algorithm we refer the interested readers to \cite{zw18,zw19}.

\subsection{Results}\label{sec:results}

To assess the quantify of the extrapolation results we used 6 metrics as introduced by \cite{sdm06} and \cite{blw06}. These figures quantify the agreement between vector fields $\mathbf{b}$ of the reconstruction and $\mathbf{B}$ of the reference model.
\begin{enumerate}
  \item[$\bullet$] vector correlation
  \begin{equation}
  C_{vec}=\displaystyle\sum_{i}\mathbf{B}_{i}\cdot \mathbf{b}_{i}/\left(\displaystyle\sum_{i}|\mathbf{B}_{i}|^2\displaystyle\sum_{i}|\mathbf{b}_{i}|^2\right)^{\frac{1}{2}},
  \end{equation}
  \item[$\bullet$] Cauchy-Schwarz inequality
  \begin{equation}
  C_{CS}=\frac{1}{N}\displaystyle\sum_{i}\frac{\mathbf{B}_{i}\cdot \mathbf{b}_{i}}{|\mathbf{B}_{i}||\mathbf{b}_{i}|},
  \label{eq:ccs}
  \end{equation}
  \item[$\bullet$] normalized vector error
  \begin{equation}
  E_{N}=\displaystyle\sum_{i}|\mathbf{B}_{i}-\mathbf{b}_{i}|/\displaystyle\sum_{i}|\mathbf{B}_{i}|,
  \end{equation}
  \item[$\bullet$] mean vector error
  \begin{equation}
  E_{M}=\frac{1}{N}\displaystyle\sum_{i}\frac{|\mathbf{B}_{i}-\mathbf{b}_{i}|}{|\mathbf{B}_{i}|},
  \label{eq:em}
  \end{equation}
  where N is the number of grid points in the computation box.
  \item[$\bullet$] field line divergence (FLD) metric: Tracing field lines from a random point on the bottom boundary in both reference model and extrapolation. If both field lines end again on the bottom boundary, a score $p_i$ can be given to this point with the distance between the two endpoints divided by the length of the field line in the reference model. Then a single score can be assigned by the fraction of the area in which $p_i$s are less than 10\%. An alternative score can be given by the fraction of flux.
\end{enumerate}

We chose a small volume for comparison to focus on the compact region and minimize boundary effects as well. The $C_{vec}$, $C_{CS}$, $E_{N}$, and $E_{M}$ were applied to the white box of Fig.~\ref{fig:ref_model} with a height of 30 Mm. The FLD metric traces field lines from each photospheric point in the green box of Fig.~\ref{fig:ref_model} (a). We note that those traced lines may pass out of the green box. Tab.~\ref{tab:merit_nue} shows results of runs with various $\nu$. We find within a reasonable range of $\nu$ the MHS extrapolation performs stably. A large $\nu$ forces the bottom boundary to agree with observation by compromising on balance of magnetic and non-magnetic forces. If $\nu$ is too small, deviation between the bottom boundary and observation can be large thus the result will be decoupled from the magnetogram. The choice $\nu=0.001$ seems the optimal and is used in the following modeling.

In the detailed resulting analysis of the test with the optimal $\nu$ we divide the domain into two height ranges, [0, 10] Mm and [10, 30] Mm. The comparison in the lower region focuses mainly on the flux rope while the higher region comparison focuses mainly on the ambient potential field. The quantitative evaluations are given in Tab.~\ref{tab:merit}. The MHS extrapolation scores better than the NLFFF extrapolation in almost all metrics in both height ranges. In the higher region, the two extrapolations perform similarly to each other. The biggest difference among all metrics is less than 3.8\%. The magnetic fields in this region consist of long and open loops which are nearly current-free. They can be well reconstructed even by a potential field extrapolation \citep{dsb09}.

\begin{figure*}
  \centering
  \includegraphics[width=17cm]{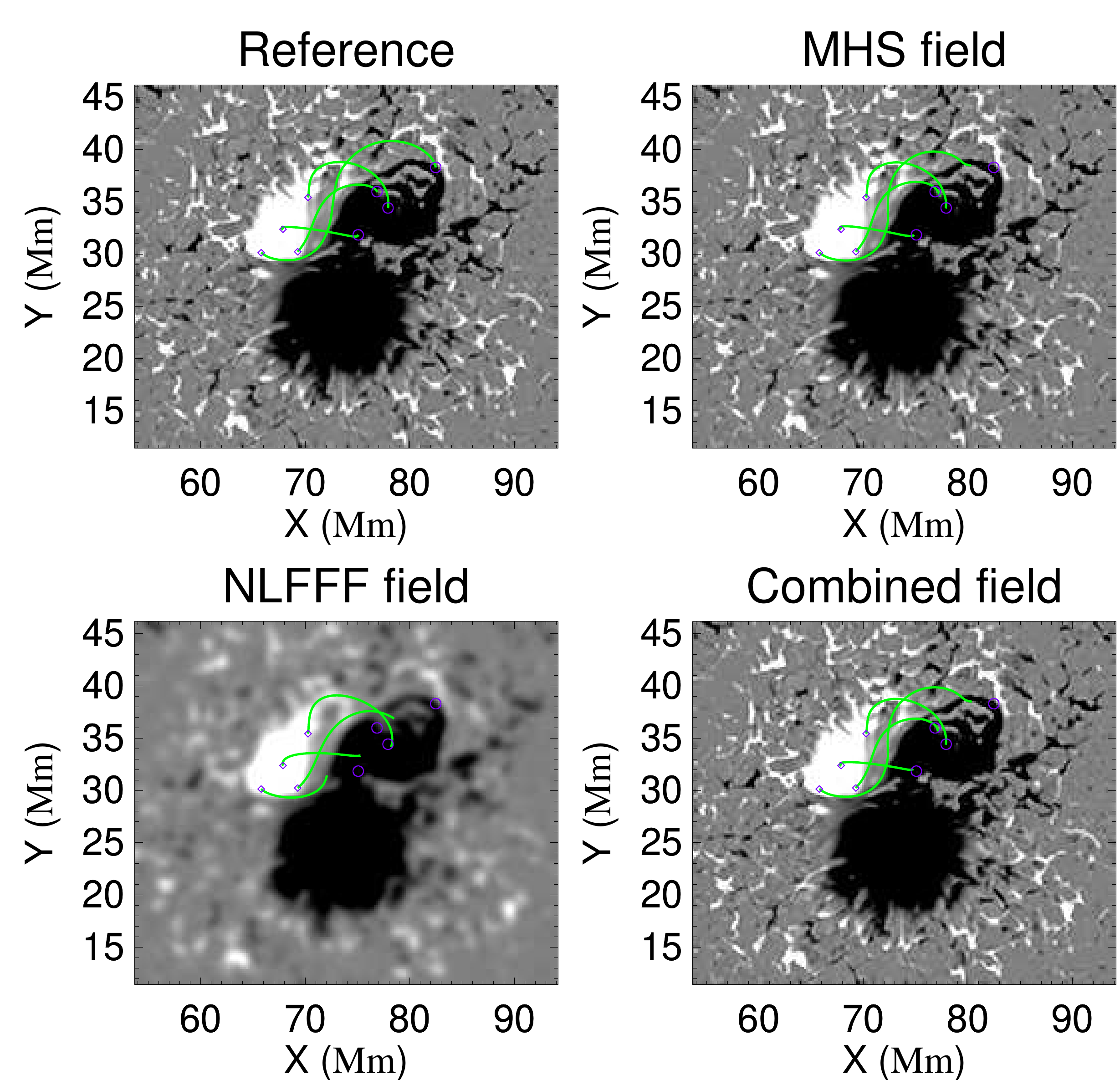}
  \caption{Magnetic field configurations of different models. The same start points marked with rhombus are selected for all panels. End points of the referenced field lines are marked with circles. Magnetogram of the NLFFF extrapolation is smoother than other panels due to preprocession.}
  \label{fig:fld_line_compare}
\end{figure*}

The comparison in the lower region, however, shows somewhat larger difference between the two extrapolations. Metrics $E_N$ and $E_M$ show that the MHS solution is 6\% better than the NLFFF solution. These two metrics are sensitive to both angle and norm differences of the magnetic field while $C_{vec}$ and $C_{CS}$ are sensitive to angle difference only. Among all metrics we used, the FLDs are the most rigorous two. The MHS extrapolation scores 0.72/0.76 in the two metrics, which are dramatic improvements ($> 40\%$) over the NLFFF's scores of 0.51/0.55. Although angel and norm differences of the magnetic field are small in general, difference of the magnetic connectivity could be significant because of the accumulation of deviation when lines are traced. In Fig.~\ref{fig:fld_line_compare} we compare the reference model with various extrapolation models. The MHS extrapolation shows better agreement with the reference model than the NLFFF extrapolation both in field lines pattern and connectivity. The magnetic field connectivity is crucial to explain eruptions and determine regions where reconnection is possible \citep{jzs19}.

The MHS extrapolation costs 14 hours on the run with Intel Xeon Gold 6150 CPU (18 cores) while the NLFFF extrapolation only costs 0.2 hour. The low convergence speed of the MHS extrapolation brings difficulties in modeling temporal evolution of the magnetic fields with high spatial resolution magnetogram (e.g. DIKST, GST).

\section{Incorporation of two extrapolations}\label{sec:mhsnlff}

The above test showed that the MHS extrapolation works well in both chromosphere and corona. However, the extremely low efficiency will obstruct its application to high spatial resolution magnetograms with large field of view. Considering that the non-force-free region is thin with a height less than 2 Mm \citep{mjm95,zwd16,zws20}, it is straightforward to apply the MHS extrapolation only to the low part (include non-force-free region plus small part of force-free region) while the force-free corona is modeled by the NLFFF extrapolation. As the MHS solution includes a consistently transition from non-force-free to force-free regions we can cut it transversely to get vector magnetogram at a force-free height. The vector magnetogram will be used as the bottom boundary for the NLFFF extrapolation. Here comes two questions. At what height shall we cut the MHS solution to get a suitable magnetogram? How thin the computational box can be set for the MHS extrapolation to get the suitable magnetogram?

\subsection{Determining the bottom boundary for the NLFFF extrapolation}

\begin{figure*}
  \centering
  \includegraphics[width=15cm]{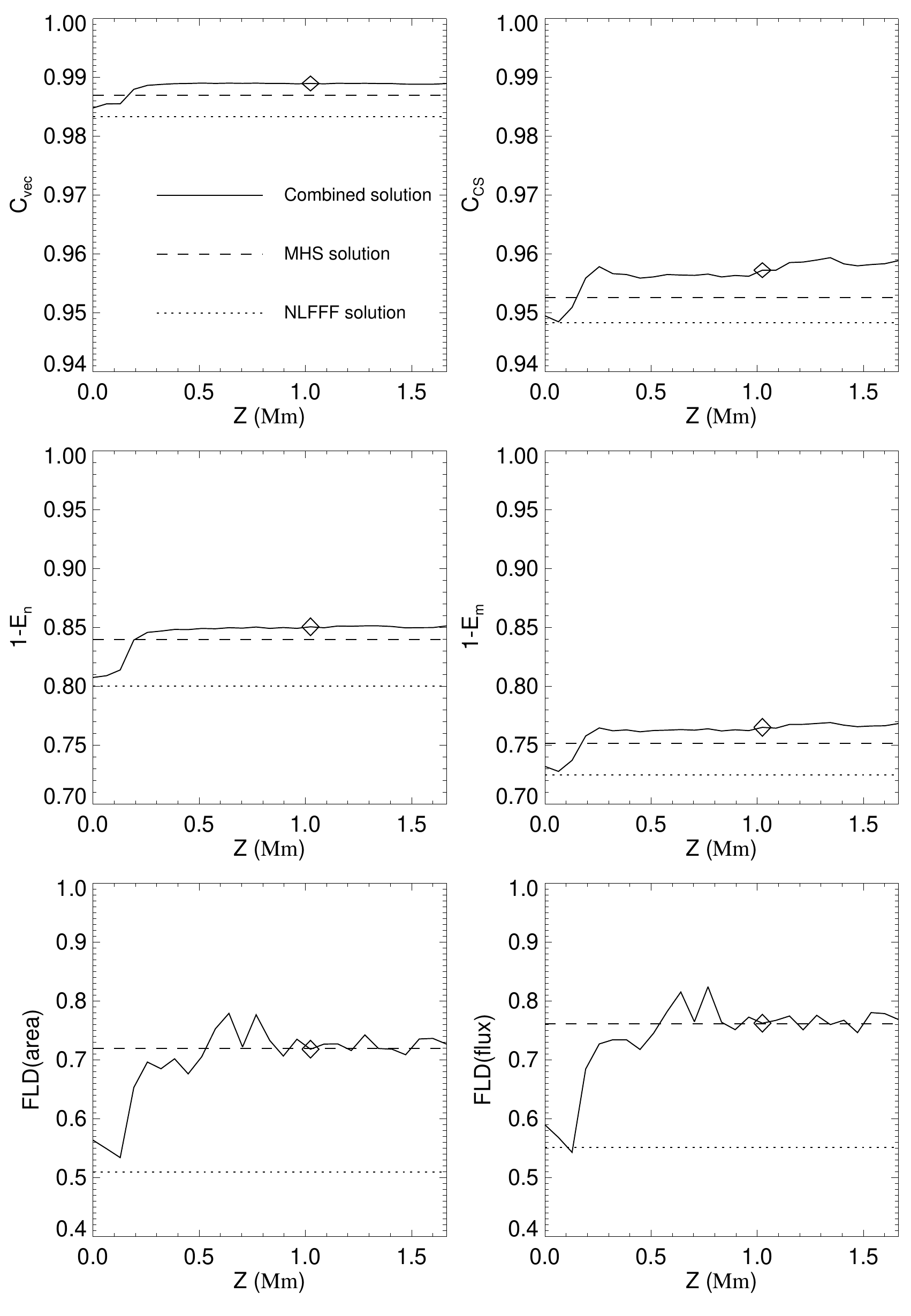}
  \caption{Figures of merit for comparison (below 30 Mm) of the reference model and the combined extrapolation with magnetograms at various heights as boundary. The dashed and dotted lines show results of the NLFFF extrapolation and the MHS extrapolation, respectively. The optimal height $1$ Mm is marked with rhombus.}
  \label{fig:heights}
\end{figure*}

The magnetic fields in the entire region do not necessarily become force-free at the same height. We can, of course, get a transverse slice of the MHS solution well in the corona to ensure its force-freeness at anywhere for a following NLFFF extrapolation. However, high slices require a relatively high computational box of the MHS extrapolation, which results in a limited improvement of computational efficiency. Thus the slice should be force-free and, at the same time, as low as possible. Fig.~\ref{fig:heights} shows how magnetograms at different heights influence extrapolation results. Poor performance can be seen when magnetograms near the bottom plane are used. As using higher magnetograms, the results improve quickly and finally converge. We consider $h=1$ Mm (marked with rhombuses) as an optimal height of the force-free vector magnetogram, above which the extrapolation shows no significant improvement.


\subsection{Determining height of computational box for the MHS extrapolation}

\begin{figure*}
  \centering
  \includegraphics[width=15cm]{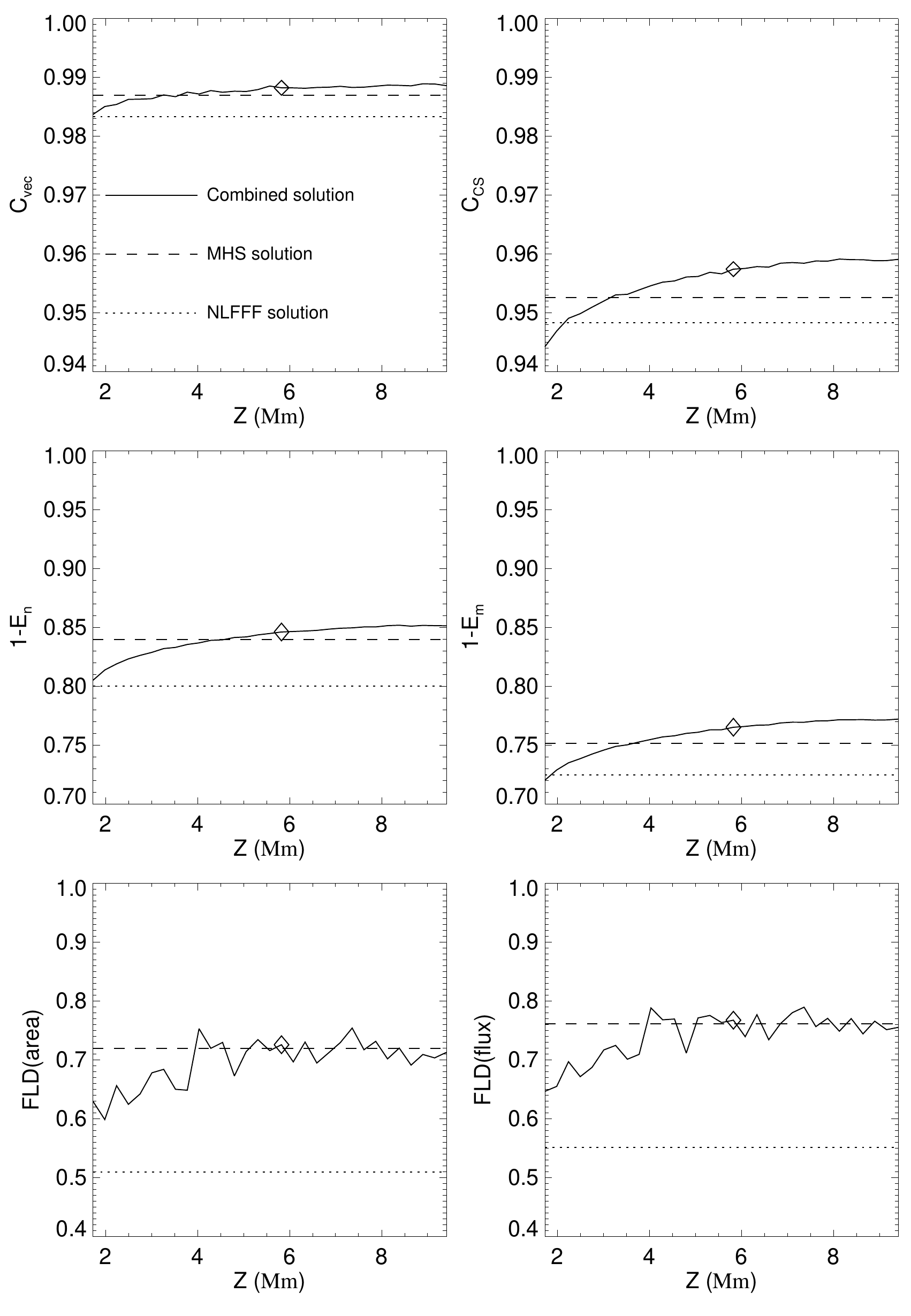}
  \caption{Figures of merit for comparison (below 30 Mm) of the reference model and the combined extrapolation with various heights of the MHS extrapolation box. The magnetogram at $1$ Mm is used as boundary input to the NLFFF extrapolation. The optimal height $5.8$ Mm is marked with rhombus.}
  \label{fig:heights2}
\end{figure*}

In the above subsection, we find through tests that a magnetogram at $h=1$ Mm is an optimal choice to link two extrapolations. Therefore the final question to be answered is how thin the computational box of the MHS extrapolation can be set. The MHS extrapolation in the above subsection is done in a $98\times49\times42$ Mm box, which is computationally expensive. In this subsection, we aim to find out the optimal height of the box for the MHS extrapolation. The box should be, at one hand low enough to reduce computation amount and, at another hand high enough to suppress upper boundary affection. Fig.~\ref{fig:heights2} shows how heights of the MHS extrapolation box influence extrapolation results. Upper boundary affection is obvious for a flat box with a height smaller than 4 Mm roughly. We consider $h=5.8$ Mm (marked with rhombuses) as an optimal height of the MHS extrapolation box. Higher box leads no significant improvement of the result. It is worth noting that the FLD indexes in both Fig.~\ref{fig:heights} and Fig.~\ref{fig:heights2} show larger oscillations than those of other indexes, which might be due to nonlinearity of the tracing process. However, the trend of change is obvious.

\subsection{Algorithm of the combined extrapolation}

The current version of our code works as follows:
\begin{enumerate}
  \item Perform a MHS extrapolation \citep{zw18,zw19} by using an observed magnetogram within a 5.8-Mm-high box.
  \item Cut the MHS solution at $h=1$ Mm to get a magnetogram.
  \item Perform a NLFFF extrapolation \citep{w04} by using the synthetic magnetogram. Notice that the synthetic chromospheric magnetogram is force-free and smooth enough to not need preprocessing.
  \item Combine the MHS solution below $1$ Mm and the NLFFF solultion together to a new solution.
\end{enumerate}
The above approach of extrapolation costs 2 hour in running the case in this study, which is 7 times faster than the original MHS extrapolation. The efficiency improvement is consistent with the choose of 5.8-Mm-high box which is roughly one seventh of the height of the original MHS extrapolation box (42 Mm).

\section{Conclusion}\label{sec:conclusion}

In this study, we have validated the MHS extrapolation with a snapshot of an RMHD simulation. Results have shown that the MHS extrapolation is more accurate than the NLFFF extrapolation to model the multiple layers on the Sun. However, it requires a much longer computational time. A new approach which combines the MHS extrapolation and the NLFFF extrapolation reaches the accuracy of the former one and, at the same time, inherits the efficiency of the later one. It is worth noting that, in the above test, we used only one snapshot (post-flare) as the reference model. To show the robustness of our method, we also selected a pre-flare snapshot (30 minutes before flare) to further test the method. We find, from the quantitative comparison showed in Tab.~\ref{tab:merit_preflare} for the additional test, that both the MHS extrapolation and the combined approach work well. Conclusions from the post-flare case test still hold in the pre-flare case.

For active region 11768 observed by SUNRISE/IMaX \citep{mda11} with spatial resolution of 100 km and pixel spacing of 40 km, a $2000\times2000\times1280$ box spans a volume of $80\times80\times51$ $Mm^{3}$ roughly contains the main magnetic structure. With above grid points setting, the NLFFF extrapolation costs 10 hours to finish, the MHS extrapolation costs 550 hours, and the combined extrapolation costs 70 hours. The combined extrapolation is moderately efficient for application to magnetograms with high resolution.





\begin{table}
  \caption{Metrics for the MHS extrapolations with various $\nu$. Comparisons are made below 30 Mm.}             
  \label{tab:merit_nue}      
  \centering                          
    \begin{tabular}{c c c c c c c}        
    \hline\hline                 
    \noalign{\smallskip}
    $\nu$ & $C_{vec}$ & $C_{CS}$ & $1-E_{n}$ & $1-E_{m}$ & $FLD (area)$ & $FLD (flux)$ \\    
    \hline
    \noalign{\smallskip}
    0.01    & 0.99 & 0.95 & 0.84 & 0.75 & 0.64 & 0.69 \\
    \noalign{\smallskip}
    0.001   & 0.99 & 0.95 & 0.84 & 0.75 & 0.73 & 0.77 \\
    \noalign{\smallskip}
    0.0001  & 0.99 & 0.95 & 0.84 & 0.75 & 0.70 & 0.75 \\
    \noalign{\smallskip}
    0.00001 & 0.99 & 0.95 & 0.82 & 0.74 & 0.57 & 0.61 \\
    \noalign{\smallskip}
    \hline                                   
  \end{tabular}
\end{table}

\begin{table}
  \caption{Metrics for the two extrapolations applied to the vector magnetogram of the flare simulation.}             
  \label{tab:merit}      
  \centering                          
    \begin{tabular}{c c c c c c c c}        
    \hline\hline                 
    \noalign{\smallskip}
    Model & $C_{vec}$ & $C_{CS}$ & $1-E_{n}$ & $1-E_{m}$ & $FLD (area)$ & $FLD (flux)$ & time (hours) \\    
    \hline                        
    \noalign{\smallskip}
    Ref.   & 1.00 & 1.00 & 1.00 & 1.00 & 1.00 & 1.00 & --\\      
    $z\le 10 Mm$\\
    \noalign{\smallskip}
    NLFFF  & 0.98 & 0.93 & 0.81 & 0.68 & 0.51 & 0.55 & 0.2\\
    \noalign{\smallskip}
    MHS    & 0.99 & 0.93 & 0.86 & 0.72 & 0.72 & 0.76 & 14\\
    \noalign{\smallskip}
    \noalign{\smallskip}
    $10\le z\le 30 Mm$\\
    \noalign{\smallskip}
    NLFFF  & 0.97 & 0.96 & 0.78 & 0.75 & -- & -- & --\\
    \noalign{\smallskip}
    MHS    & 0.98 & 0.96 & 0.81 & 0.77 & -- & -- & --\\
    \noalign{\smallskip}
    \hline                                   
  \end{tabular}
\end{table}

\begin{table}
  \caption{Metrics for extrapolations with pre-flare reference model. Comparisons are made below 30 Mm.}             
  \label{tab:merit_preflare}      
  \centering                          
    \begin{tabular}{c c c c c c c}        
    \hline\hline                 
    \noalign{\smallskip}
    Model & $C_{vec}$ & $C_{CS}$ & $1-E_{n}$ & $1-E_{m}$ & $FLD (area)$ & $FLD (flux)$ \\    
    \hline
    \noalign{\smallskip}
    NLFFF    & 0.98 & 0.97 & 0.81 & 0.78 & 0.44 & 0.44 \\
    \noalign{\smallskip}
    MHS      & 0.99 & 0.97 & 0.85 & 0.80 & 0.62 & 0.65 \\
    \noalign{\smallskip}
    Comb.    & 0.99 & 0.97 & 0.85 & 0.81 & 0.68 & 0.72 \\
    \noalign{\smallskip}
    \hline                                   
  \end{tabular}
\end{table}

\clearpage

\begin{acknowledgements}
This work is supported by the mobility program (M-0068) of the Sino-German Science Center. TW acknowledges DLR-grant 50 OC 2101.
\end{acknowledgements}

%
   \bibliographystyle{aa} 
   \bibliography{aa} 
%

\end{document}